# Network Intrusion Forensic Detection System with Explainable Artificial Intelligence and Digital Chain of Custody

José Luis Vela[1]*, Carmen Pellicer[1]
[1]CESTE, University Center, Zaragoza, Spain
*Corresponding author: José Luis Vela Alonso (jlvela@ceste.com)

**Highlights**

- Dual-mode architecture certifies models and freezes them for forensic use.
- OCSF normalization resolves feature-coverage gaps across heterogeneous evidence.
- Causal flow aggregation raises multiclass F1-macro from 0.39 to 0.93.
- A hidden binary-only certification bug is found via a real packet capture.
- SHAP-based KPIs operationalise explainability and traceability for admissibility.

## Abstract

Deploying an explainable intrusion detection model from a validated laboratory setting into an operational forensic workflow requires guarantees that laboratory benchmarks do not test, namely that the certified model is the exact artifact applied to a case, that heterogeneous evidence can be classified reliably, and that every step is verifiable. This paper presents MODEXRE, a forensic pipeline that extends a previously validated synthetic-data-trained XGBoost classifier with SHAP explainability into a dual-mode architecture, Laboratory and Formal, governed by a verifiable chain of custody and a frozen, hash-checked model. The separation also lets models be trained on public and synthetic data while private evidence stays confined to Formal mode.

Four design decisions, each triggered by an observed failure rather than anticipated in advance, shaped the final architecture: an information leak from a timestamp artifact inflated Kitsune performance before correction (F1-macro 0.984 to 0.662, regression-tested); a coverage gap between the 78-feature reference model and the fields real evidence can populate caused a real SSH alert to be misclassified, resolved with a six-feature representation; causal temporal aggregation raised Kitsune multiclass F1-macro from 0.394 to 0.932; and a real packet capture revealed that the certification pipeline had always trained a binary classifier, moving from classifying 100% of legitimate traffic as malicious to about 87% correct Normal classification after correction.

The default four-class certified model (v1) achieves F1-macro = 0.9538 on CICIDS2017-derived evidence, and the recertified 18-category model (v3) reaches F1-macro = 0.6214, a figure that reflects the real multiclass task once the certification defect was corrected, with specialized port-scan detectors reaching F1-macro up to 0.9997. SHAP attributions are consistent with domain knowledge, and forensic KPIs, chain-of-custody verification, model integrity hashing, and per-prediction explainability, are reported as measured outcomes.

## 1. Introduction

No prior framework in the intrusion-detection literature reports the transition of a laboratory-validated, explainable classifier into a certified forensic pipeline together with the failures that transition exposes; this paper reports both the pipeline and the failures. A model that performs well on a held-out test split provides no guarantee that the same model, applied months later to a real case, is the artifact that was originally validated, nor that the heterogeneous, partially populated evidence encountered in the field resembles the curated features it was trained on (Carrier & Spafford, 2003; Casey, 2011; Garfinkel, 2010; Pollitt, 2010). Forensic deployment introduces requirements that benchmark evaluation does not test: model identity over time, feature availability across incompatible evidence sources, and detectability of the system's own failure modes before they reach a case report.

A companion paper (Vela Alonso & Pellicer, 2026) validated a synthetic-data-trained XGBoost classifier (Chen & Guestrin, 2016) with SHAP explainability (Lundberg & Lee, 2017) under the Train-on-Synthetic, Test-on-Real protocol, achieving F1-macro = 0.96 on CICIDS2017 while preserving evidentiary separation between original data and derived artifacts. That work resolved the data-availability and transparency barriers to forensic adoption of machine learning-based intrusion detection. It did not, however, address what happens once a certified model leaves the laboratory: whether the exact validated artifact can be guaranteed unchanged when applied to a case, whether real evidence sources (intrusion detection alerts, firewall logs, raw packet captures) can populate the feature vector the model expects, and whether attack patterns that only manifest across sequences of events, such as port scanning, can be detected at all under an event-by-event classification scheme.

This paper addresses that gap through MODEXRE (MOdular Digital EXamination and Reporting Engine), a forensic pipeline built around a single architectural principle: strict separation between a Laboratory mode, where models are trained, tuned, and certified without chain-of-custody obligations, and a formal mode, where a frozen, hash-verified certified model performs inference exclusively on real evidence under full chain-of-custody logging. This separation operationalizes, as a structural property of the architecture rather than a downstream audit, the requirement identified across the forensic admissibility literature that an artificial intelligence system's output can be tied unambiguously to a specific, unaltered, previously validated model (Section 2.3). The separation also resolves a practical constraint: case evidence is scarce and private, so training certified models on public and synthetic data in Laboratory mode, with real evidence confined to Formal mode, allows more powerful models than any trained ad hoc on a single case, through a fully auditable training process.

Within this architecture, four design decisions emerged during development, each triggered by an empirically observed failure rather than anticipated in advance. First, an absolute-timestamp field in the Kitsune dataset produced an artificially inflated classifier that had learned dataset construction artifacts rather than genuine traffic patterns. Second, a coverage gap between the 78-feature academic reference model and the handful of fields real evidence sources can reliably populate caused a real attack to be misclassified as benign. Third, event-level classification proved structurally incapable of detecting attacks, port scanning foremost among them whose signature exists only in the temporal relationship between multiple flows from the same origin. Fourth, and most consequential, analysis of a real packet capture revealed that the certification pipeline had silently trained a binary classifier throughout development, regardless of the multiclass labels available in the training data. Each decision is reported with the problem that triggered it, the root-cause analysis that identified it, the correction applied, and the regression test that now protects against its recurrence.

The key contributions of this work are:

- **Dual-mode certified architecture:** a Laboratory / Formal separation that ties every Formal-mode inference to a specific, hash-verified, previously certified model, operationalizing model-identity guarantees identified as necessary but rarely implemented for forensic AI admissibility. By decoupling training data from case evidence, it also enables more powerful, auditable models than case-specific training could provide.
- **Multi-source evidence normalization via OCSF:** a common representation across Suricata alerts, CEF firewall logs, and raw packet captures that expose, and then resolve the feature-coverage gap between academic reference models and real evidence.
- **Causal temporal flow aggregation:** a real-time-compatible aggregation layer that raises multiclass F1-macro on Kitsune from 0.394 to 0.932 by making crossflow behavioral patterns visible to a per-event classifier.
- **Empirical discovery and correction of a hidden certification defect:** a binary-only training bug, invisible on every benchmark dataset used during development, exposed only by evaluating the system against a real packet capture before and after correction, the most direct empirical validation reported in this work.
- **Auditable forensic KPIs:** chain-of-custody integrity verification, certified-model hash checking, and per-prediction SHAP explainability, reported as measured outcomes rather than design intentions.

The remainder of this paper is structured as follows. Section 2 reviews related literature on digital chain of custody, cybersecurity event normalization, AI admissibility as forensic evidence, and temporal flow aggregation for sequence-dependent attack detection. Section 3 details the forensic pipeline architecture. Section 4 presents experimental evaluation of the four design decisions and the final certified model. Section 5 presents a port-scanning case study with SHAP-based explainability. Section 6 discusses forensic implications, regulatory alignment, and limitations. Section 7 concludes.

## 2. Related Work

### *2.1 Digital chain of custody and evidence integrity*

Traditional forensic chain of custody documents who accessed a piece of evidence, when, and for what purpose, from collection to courtroom presentation (ENFSI, 2015). Digital evidence complicates this model: unlike physical evidence, it can be copied, moved, or altered without leaving a perceptible trace unless a mechanism exists to prevent or detect it. The literature agrees that documentation alone, without a cryptographic verification mechanism, cannot guarantee evidentiary integrity (Cosić & Bača, 2011).

Hash-chaining, where each evidence element's fingerprint incorporates the hash of the preceding element so that alteration of any element invalidates every subsequent fingerprint, remains the most widely adopted mechanism given its favorable cost-to-guarantee ratio (Bradford & Ray, 2007). A parallel line of work explores distributed-ledger approaches, principally blockchain, to remove reliance on a single custodian who could otherwise alter the integrity record undetected, ranging from fuzzy-hash schemes tolerant of legitimate evidence alteration to permissioned-blockchain evidence management frameworks that let the defense challenge evidentiary fidelity in court (Lone & Mir, 2019).

This literature suggests a practical hierarchy of guaranteed levels. A first level relies on access-metadata logging alone, without cryptographic integrity verification. A second level adds hash-chaining, making alteration mathematically detectable, though the hash registry itself remains a single system susceptible to administrative-level tampering. A third level would add digital signatures under a public-key infrastructure, enabling non-repudiable attribution of responsibility, not merely detection of alteration.

A fourth level substitutes or supplements the custody registry with a distributed ledger, removing single-custodian trust entirely. Systematic reviews of blockchain-based custody proposals report that many do not progress beyond laboratory prototyping, given the mismatch between implementation complexity and marginal benefit when the system's own guardian is also the party who must prove its integrity to a third party.

Even in its simplest form, hash-chaining already satisfies part of what a court requires to validate evidentiary reliability: it demonstrates that an evidentiary element has not been altered since its fingerprint was generated, and it detects without necessarily attributing any subsequent tampering attempt (Bharati, 2025). The distinction between detecting alteration and attributing responsibility for it is central to the guaranteed level selected for MODEXRE's chain of custody (Section 3.5).

### *2.2 The OCSF standard for cybersecurity event normalization*

A persistent practical obstacle in network forensic analysis is the heterogeneity of formats in which security systems (intrusion detection systems, firewalls, endpoint protection, mail platforms) report events, historically forcing the construction of source-specific parsers and hindering both cross-tool event correlation and the reuse of a single classification model across evidence sources of different origin (OCSF Consortium, 2024). The Open Cybersecurity Schema Framework (OCSF), an open standard initially backed by a consortium including AWS, Splunk, and IBM and later adopted by the Linux Foundation, defines a taxonomy of common security event classes with normalized attributes, so that a network activity event, a detection alert, or an authentication log are represented coherently regardless of the product that generated them.

Two properties of the standard are directly relevant here. First, its agnosticism toward serialization format (accepting JSON, Parquet, or Avro) makes it suitable as an intermediate normalization layer between disparate evidence sources and a homogeneous feature-extraction pipeline. Second, the standard explicitly acknowledges that not every source can populate every attribute of an event class (attributes are divided into required, recommended, and optional precisely to accommodate data producers of varying granularity), a conceptual framework directly applicable to the feature-coverage problem addressed in this paper (Section 4.3).

***2.3 Admissibility of artificial intelligence systems as forensic evidence***

The literature identifies a structural tension in applying evidentiary reliability standards to black-box machine learning systems: the opacity of high-parameter-count models means that not even their own developers can fully explain how a given output was produced, which conflicts directly with the requirement for a known error rate and an explainable methodology (AI Legal Authority, 2025). Multiple sources converge on the position that explainability is not a desirable property but a necessary condition for an AI system to clear the admissibility threshold: because a court must assess system reliability, it must be possible to explain how a given conclusion was reached and with what degree of certainty (Doshi-Velez & Kim, 2017; Justice Speakers Institute, 2025).

This regulatory pressure is not confined to a single jurisdiction. In the United States, a new Federal Rule of Evidence, Rule 707, was proposed in November 2024 and approved by the relevant committee on 10 June 2025, requiring AI-generated evidence without accompanying human expert testimony to meet a Daubert-equivalent reliability standard rather than benefit from a presumption of reliability based on its apparently objective, computational origin (Rumberger Kirk, 2025).

The European Union's AI Act independently converges on the same requirement through regulatory rather than evidentiary channels, classifying AI systems deployed by law enforcement authorities or within judicial proceedings as high-risk and imposing reinforced obligations of transparency, technical documentation, and human oversight. A network intrusion detection system whose outputs may be used in disciplinary or judicial proceedings falls within this risk category.

The explainability and traceability requirements the U.S. literature frames in evidentiary terms thus find, in the European framework, their counterpart in regulatory compliance terms, a double convergence that reinforces the decision to model explainability and traceability as architectural requirements of MODEXRE rather than functionality added after the fact.

### *2.4 Temporal flow aggregation for sequence-dependent attack detection*

A model trained on the features of an individual flow can correctly identify whether that specific flow is anomalous, but attacks such as port scanning manifest precisely in the relationship between many short flows directed from the same origin toward many destinations or ports, a relationship invisible when each flow is evaluated in isolation. Flow-based scan-detection literature addresses this by fragmenting the continuous traffic stream into fixed-size temporal windows and computing, per window and per source address, derived attributes such as the number of distinct destinations addresses or ports contacted.

Ring, Landes, and Hotho (2018) capture this aggregated scan behavior and compare an unsupervised approach, based on sequential hypothesis testing, against a supervised classification approach, concluding that supervised approaches achieve better detection rates at the cost of requiring labeled data, generally harder to obtain in real environments than in a laboratory dataset. A complementary line of work highlights that the choice of aggregation level individual flow versus aggregated flow directly affects which attack families are detectable and at what performance, recommending per-attack-type evaluation rather than a single global performance metric (Rodríguez et al., 2022), a conclusion directly applicable to MODEXRE's design (Section 3.6).

### *2.5 Synthesis and research gap*

The reviewed literature generates well-developed guarantee frameworks for digital chain of custody, largely independent of intrusion-detection systems; OCSF solves the event-normalization problem but not the question of how a model trained on laboratory-level feature coverage should be adapted to unequal coverage across real evidence sources; legal literature on AI admissibility explicitly demands explainability and traceability but does not descend to the level of specific design decisions that satisfy that demand; and flow-aggregation literature solves the technical detection problem without integrating it into a complete forensic pipeline with end-to-end integrity guarantees.

MODEXRE sits at the intersection of these four bodies of knowledge, and its contribution is not a novel technique within any single one, but their integration into a single operational architecture: an explainable, frozen, and versioned classification model; an OCSF-based normalization layer that absorbs source heterogeneity; a verifiable chain of custody accompanying every operation; and a temporal aggregation stage that makes sequence-dependent attack patterns visible.

## 3. Methodology

### *3.1 Design principles*

MODEXRE follows a constructive applied-research approach: rather than testing a hypothesis through primary data collection, it builds, implements, and validates a software artifact that resolves an engineering problem with forensic and legal implications, evaluated against both quantitative model-performance criteria and qualitative traceability and explainability criteria.

Two data sources are used throughout. Secondary data the CICIDS2017, UNSW-NB15, and Kitsune datasets, also used in the companion paper are reused to train and validate the classification model in Laboratory mode. Evidence generated or collected specifically for this work Suricata alerts, CEF firewall logs, and raw packet captures validate system behavior in Formal mode against evidence structurally distinct from the model's training data. This separation between training data and operational validation data is one of the central axes of MODEXRE's methodological design.

Evaluation combines quantitative and qualitative criteria deliberately, to reflect the system's dual technical and evidentiary nature. The quantitative dimension relies on multiclass classification metrics, principally F1-macro, given the imbalanced nature of network traffic categories, complemented by precision and recall when analysis concerns a specific category such as port scanning.

The qualitative dimension evaluates aspects that do not reduce to a single figure: the coherence between SHAP attributions and domain knowledge of each attack type, the system's capacity to detect and document its own limitations (illustrated by the design decisions reported in Sections 4 and 5), and the empirically verified effectiveness of the chain of custody against simulated tampering attempts in the automated test suite. Neither dimension alone is sufficient: a model with excellent metrics but incoherent explanations, or a chain of custody that is assumed but not empirically verified, does not satisfy this project's objectives.

### *3.2 Dual-mode architecture: Laboratory and Formal*

The most consequential design decision in the entire development is the separation, at both code and guarantee level, of two fully distinct operating modes (Fig. 1). Laboratory mode is the environment in which the model is trained, tuned, and validated: it mixes real data with synthetic data generated via CTGAN and the SDV library, directly inheriting the methodology validated in the companion paper; it requires no chain of custody, since it does not operate on real-case evidence; and it permits unrestricted modification of hyperparameters and training data. Model certification also occurs in this mode.

Separation also answers a practical constraint that affects any machine-learning forensic system: the evidence it must analyze is, by nature, scarce and private. Network traffic from a real case belongs to the affected organization, contains personal and business data, cannot be published or reused across investigations, and at best provides a few thousand events from a single incident; a model trained solely on each case's evidence would be ad hoc, small sample, biased towards that incident, and impossible to validate independently. Laboratory mode resolves this by decoupling the origin of training data from the origin of the analyzed evidence: training on public research datasets and on synthetic data generated with CTGAN and SDV gives access to a volume and diversity of scenarios unattainable with real evidence, without handling any case's private data. The certified model reaching Formal mode is therefore more powerful than any model trainable ad hoc, and its training process is fully auditable because it depends on no data that cannot be shown. The mode separation thus turns a limitation, the scarcity and confidentiality of evidence, into a design property.

Formal mode, by contrast, is the environment in which a previously trained and validated model is applied exclusively to the evidence underlying a specific case: it requires a mandatory, step-verifiable chain of custody, and the model to which Formal mode is applied is frozen: Formal mode can only perform inference on it, never retrain it. This separation directly answers the model-drift problem: certifying the model in Laboratory mode through versioning and an integrity hash that unambiguously identifies the exact weights and configuration of the trained model, and preventing any modification of that model within Formal mode, makes it possible to document verifiably that the model which classified a case's evidence is exactly the one validated during laboratory certification, without retraining, fine-tuning, or any change between the two moments. This guarantee, identified in Section 2.3 as a practically necessary condition for AI admissibility as forensic evidence, is thus a structural property of the architecture rather than a downstream check.

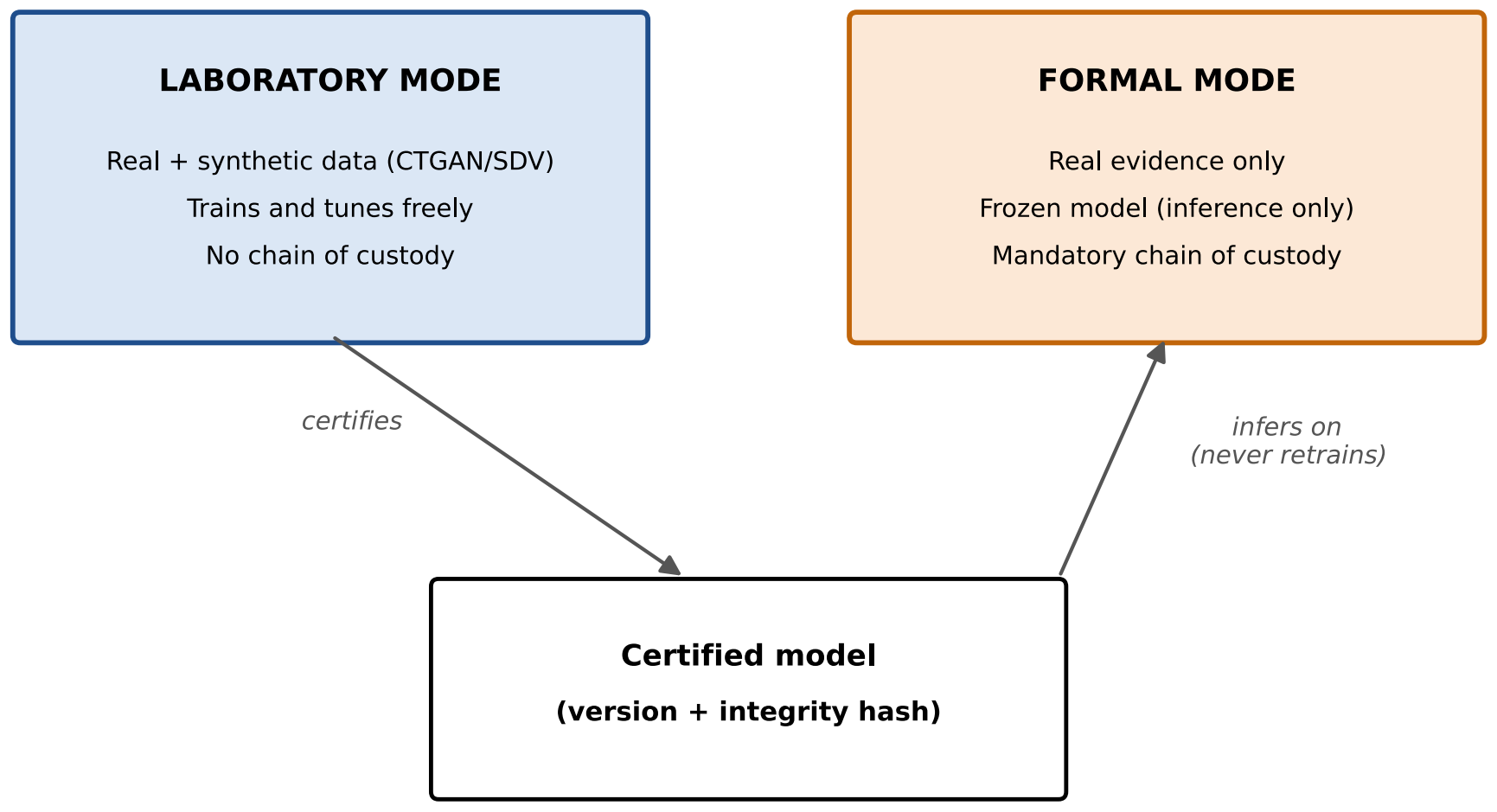

**Fig. 1.** Dual-mode architecture: Laboratory certification and Formal inference on a frozen certified model.

At implementation level, this separation is reflected in the backend's modular organization: custody/ (chain of custody), ingestion/ (evidence ingestion), ocsf/ (OCSF normalization), features/ (feature extraction), models/ (training, certification, inference), pipeline/ (formal pipeline orchestration), and report/ (forensic report generation). A Streamlit interface exposes this separation as two isolated tabs (Fig. 2), exposing only the actions valid for each mode.

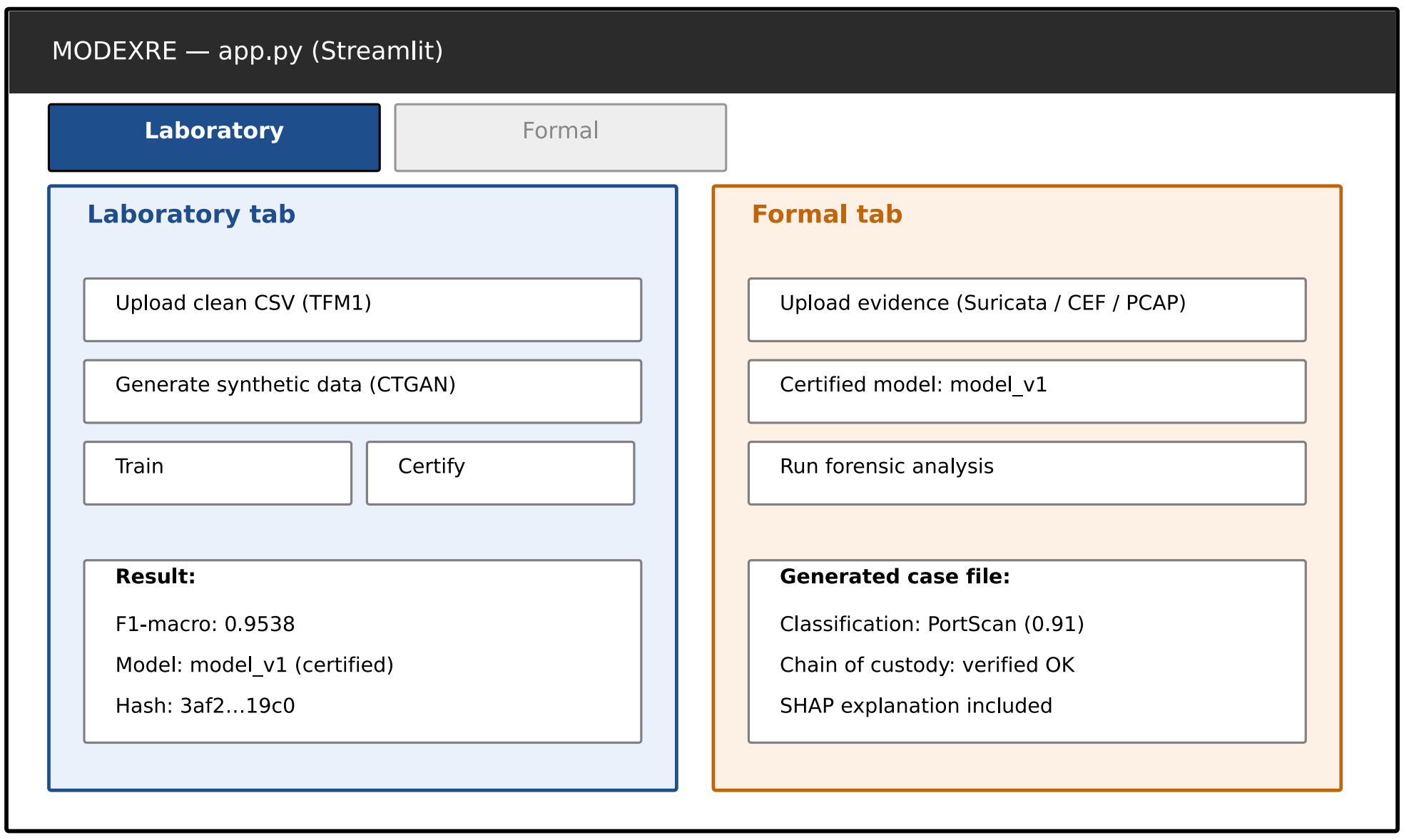

**Fig. 2.** Streamlit interface separating Laboratory and Formal workflows.

### *3.3 Attack taxonomy and evidence coverage*

MODEXRE inherits from the companion work a closed thirteen-category attack taxonomy (Normal, Fuzzers, Exploits, DoS, Reconnaissance, Generic, Analysis, Shellcode, Backdoors, DDoS, PortScan, MitM, and BruteForce), implemented in the OCSF mapping layer. Any event whose source-reported attack category falls outside this closed taxonomy is explicitly rejected rather than assigned an ad hoc category, guaranteeing that system output remains comparable across executions and evidence sources. Within this taxonomy, MODEXRE distinguishes two OCSF event classes with distinct semantics: Network Activity, representing already-aggregated flow traffic (whether pre-labeled training data or unlabeled real evidence produced by MODEXRE's own PCAP aggregation), and Detection Finding, representing a third-party system's (Suricata's or a firewall's) report on that same traffic. The latter class is never treated as ground truth: the category an IDS or firewall assigns is one contextual evidence element among others, while the final forensic classification is always and exclusively produced by MODEXRE's own model, preserving analytical independence from the tool that generated the original alert.

### *3.4 Multi-source ingestion and the coverage gap*

MODEXRE supports four evidence source types with markedly different structuring and labelling levels. CICIDS2017, UNSW-NB15, and Kitsune, in the clean CSV format inherited from the companion project, are labeled and used for Laboratory-mode training. Suricata alerts (EVE JSON), firewall logs (CEF/syslog), and raw packet captures (PCAP) are unlabeled and constitute Formal-mode evidence.

Each source has its own ingestion path: CSVs are read through direct tabular connectors; Suricata alerts are loaded line-by-line as structured JSON; CEF logs are parsed against the standard's common-core fields (source, destination, ports, protocol, action, category, message), without full coverage of vendor-proprietary dialects a deliberately documented scope limitation; and PCAP captures are processed by a purpose-built aggregator that reconstructs flows from the source/destination/port/protocol five-tuple with inactivity-based cutoff, without matching the coverage of a specialized tool such as CICFlowMeter.

This heterogeneity exposes a coverage gap central to this paper's findings: the academic reference model, trained on the complete CICIDS2017 feature set of 78 CICFlowMeter-derived attributes (Sharafaldin, Lashkari, & Ghorbani, 2018), cannot be reliably populated by any real evidence source MODEXRE supports a typical Suricata alert reliably populates only 4 of those 78 fields, with the remainder defaulting to zero. This is not an abstract concern: Section 4.3 reports the concrete forensic consequence.

### *3.5 Chain-of-custody design*

MODEXRE implements chain-of-custody guarantees at Level 2 of the hierarchy described in Section 2.1: a SHA-256-based hash chain linking every pipeline operation (ingestion, normalization, feature extraction, inference, report generation) as input-hash/output-hash pairs, chained to the preceding operation, without digital signatures. This choice, over more sophisticated public-key-infrastructure or blockchain-based alternatives identified in the literature, reflects an explicit methodological decision: prioritizing demonstration of the complete integration's viability normalization, aggregation, explainable classification, and traceability over maximizing the robustness of any single component at the cost of the operational complexity that would entail.

### *3.6 The Formal pipeline and temporal aggregation*

The Formal pipeline orchestrates, under continuous chain-of-custody supervision, the sequence: ingestion, OCSF normalization, causal temporal aggregation, feature extraction, certified-model inference, and report generation. The aggregation stage computes, for each event, three variables (distinct destination ports contacted, distinct destination hosts contacted, and number of events from the same origin) within a configurable temporal window (60 seconds by default), strictly causally: only past or simultaneous events are considered, never future ones, a restriction that could be relaxed in a laboratory setting where all evidence is known in advance but is deliberately retained here because it is the only design compatible with a system intended to operate, or potentially operate, on a live evidence stream (Fig. 3).

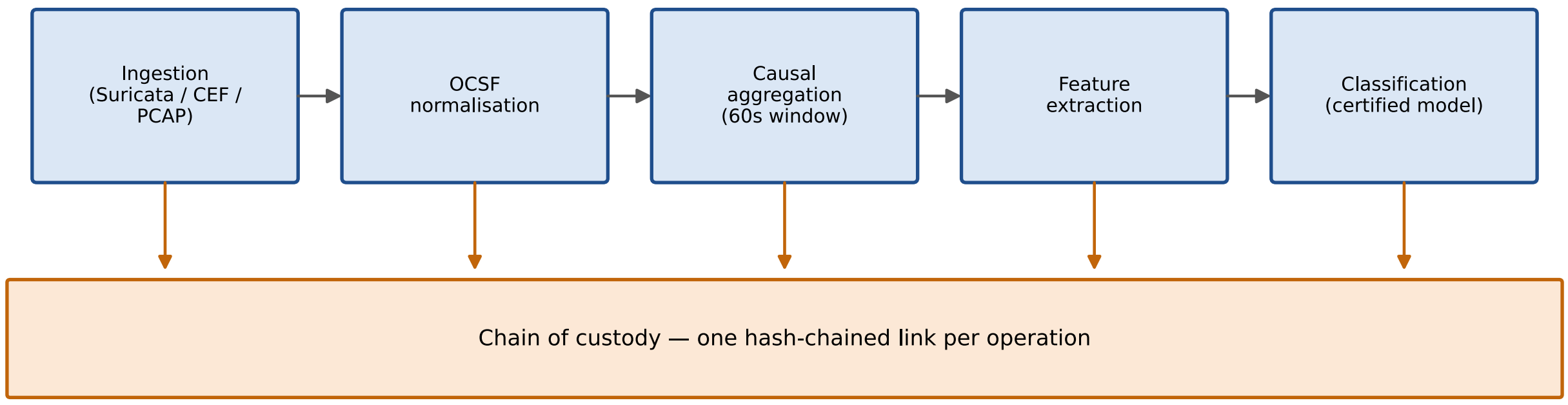

**Fig. 3.** Formal pipeline stages and their link to the chain of custody.

### *3.7 Validation through automated testing*

System validation relies on a battery of 139 automated tests distributed across twenty modules, covering each architectural layer in isolation (ingestion connectors, OCSF mappings, feature extraction, chain of custody, classifier) together with several complete end-to-end runs replicating both the Laboratory workflow (train, freeze, infer, explain) and the complete Formal workflow (ingest, normalize, extract features, classify, verify custody, generate report).

Critically, this suite includes regression tests that permanently encode the corrections derived from the design decisions reported in Sections 4 and 5 (exclusion of absolute-timestamp fields, correct behavior of causal flow aggregation, multiclass-only model certification), so that any future modification reintroducing one of these problems is caught automatically before a new model is certified.

## 4. Experimental evaluation

### *4.1 Experimental configuration*

All models use XGBoost with the following hyperparameters: n_estimators = 200, max_depth = 6, random_state = 42, logloss (binary) or mlogloss (multiclass) as the internal evaluation metric, and an 80%/20% stratified train/test split. Reported metrics are accuracy, F1-macro, F1-weighted, precision-macro, and recall-macro. The final certified Formal-mode model is trained and evaluated at 4,310 events (n_train = 3,448; n_test = 862).

### *4.2 Design decision 1: information leak in the Kitsune dataset*

An initial model trained on the only columns available port, IP, and timestamp achieved a surprisingly high F1-macro of 0.984 for so few features. SHAP analysis revealed that the highest-weighted variable was the absolute event timestamp: Kitsune concatenates captures from separate sessions, one per attack type, so the model had learned which capture session a flow belonged to rather than any genuine malicious traffic pattern information that is trivially predictable but forensically worthless.

Excluding the absolute timestamp from the feature vector reduced performance to F1-macro = 0.662, a lower but honest figure reflecting the model's real capacity to generalise from traffic patterns rather than dataset-construction artifacts. The correction was permanently incorporated into the feature extractor and protected by an automated regression test that fails if any absolute-timestamp field is reintroduced into the training vector. As a control, UNSW-NB15, which also contains absolute-time columns, was confirmed not to exhibit this problem, indicating the leak is specific to Kitsune's construction rather than a generic artifact-extraction issue.

### *4.3 Design decision 2: feature-coverage gap across evidence sources*

The reference model, trained on the complete CICIDS2017 dataset, uses 78 CICFlowMeter-derived features (Sharafaldin et al., 2018). No real evidence source MODEXRE supports can populate this vector: a Suricata alert reliably fills only 4 of the 78 variables, with the remainder defaulting to zero, causing a real SSH brute-force alert to be classified as normal traffic. A first reduced model, restricted to five reliably populated features (duration, inbound/outbound packets, inbound/outbound bytes), was certified in response (Table 1).

**Table 1.** Reference model versus first reduced model, both certified on CICIDS2017.

| Model | Number of features | F1-macro | Behavior on the SSH alert |
|---|---|---|---|
| Full (academic reference) | 78 | 0.9992 | Classifies as Normal (real coverage ≈ 5%) |
| Reduced (Formal-mode default) | 5 | 0.9655 | Correctly classifies as malicious (prob. 0.74) |

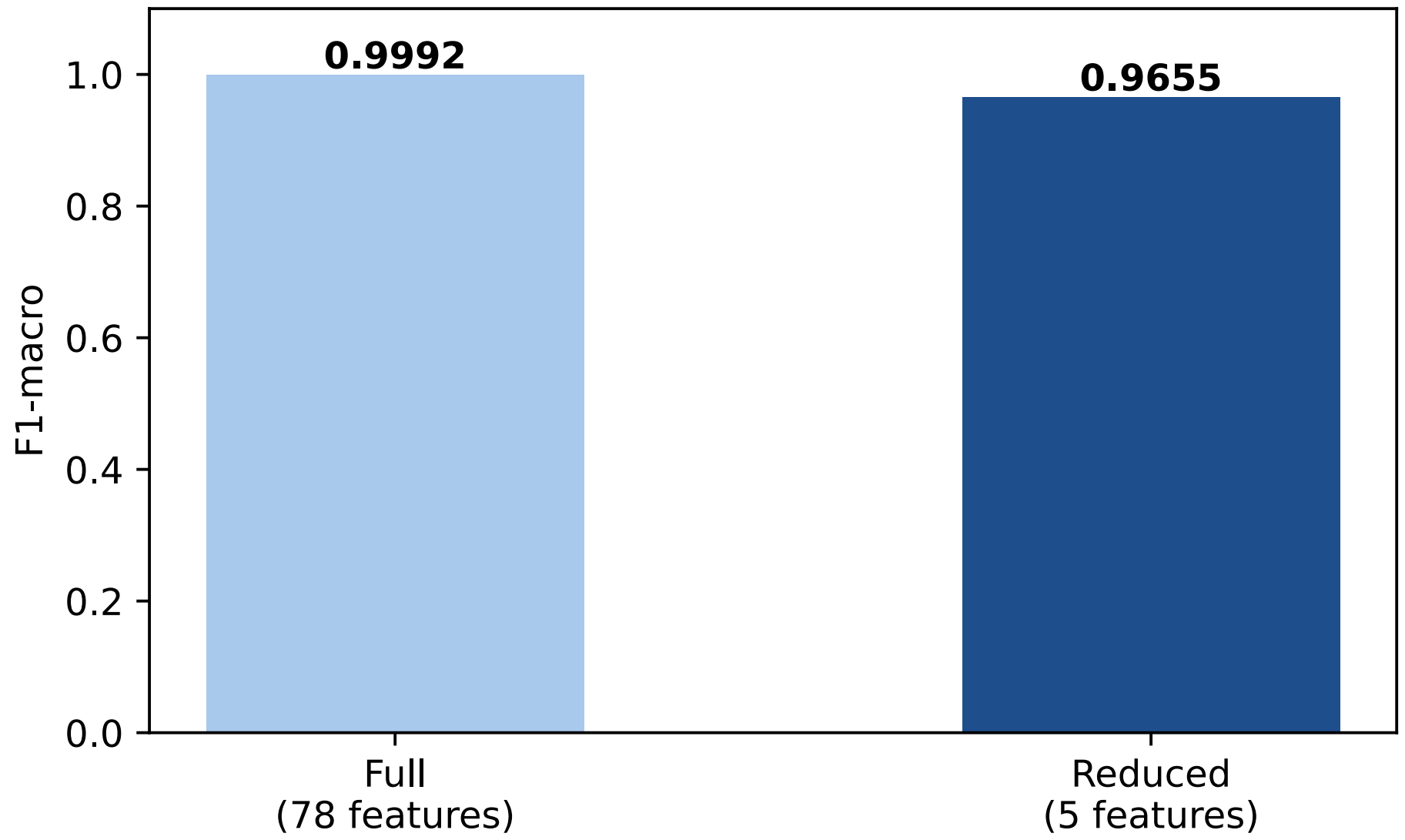

**Fig. 4.** Reference model versus reduced model performance on CICIDS2017.

The reduced model, with sixteen times fewer features, achieves F1-macro = 0.9655 against the full model's 0.9992 (Fig. 4), an expected and acceptable degradation, offset by its ability to correctly classify real evidence the full model cannot. SHAP analysis of the reduced model identified bytes_in and duration as the most influential variables in correctly classifying the SSH alert. This five-feature model was later superseded by the aggregation-enhanced version described in Section 4.4, which constitutes the final certified model (v1) reported in Section 4.5.

The implication extends beyond the raw performance figure: an F1-macro of 0.9992 that in practice classifies a genuine alert as harmless due to insufficient feature coverage is not a model with room for improvement, but one whose metric is actively misleading with respect to evidence structurally different from its training distribution, precisely the risk the black-box literature associates with opaque classifiers (Section 2.3).

### *4.4 Design decision 3: temporal aggregation for scan detection*

A specialized scan detector (Section 5), applied to a real case (a Suricata «ET SCAN SSH BruteForce» alert, a single 500-packet flow), classified it as non-scan with 0.9999 confidence, while the generic multiclass detector marked it as malicious (0.74). The specialized model was not wrong: a 500-packet flow resembles brute-force behavior more than scanning, which manifests as many short connections to distinct destinations from the same origin a pattern the pipeline, until then, could not see when classifying each event in isolation.

The correction introduced a temporal aggregation stage, between OCSF normalization and feature extraction, computing three per-event variables (distinct destination ports, distinct destination hosts, events from the same origin) within a configurable window (60 seconds by default), strictly causally (Table 2).

**Table 2.** Validation of the causal aggregation signal against a scan pattern and a contrasting case.

| Validation scenario | agg_distinct_dst_ports | Interpretation |
|---|---|---|
| 8 alerts, same origin, 8 distinct ports (last event of sequence) | 8 | Scan pattern correctly detected |
| 8 alerts, same origin, 8 distinct ports (first event of sequence) | 1 | Correct: sees only itself (causal computation) |
| 5 events to the same port (repeated normal traffic) | 1 | Correctly distinguished from scan pattern |

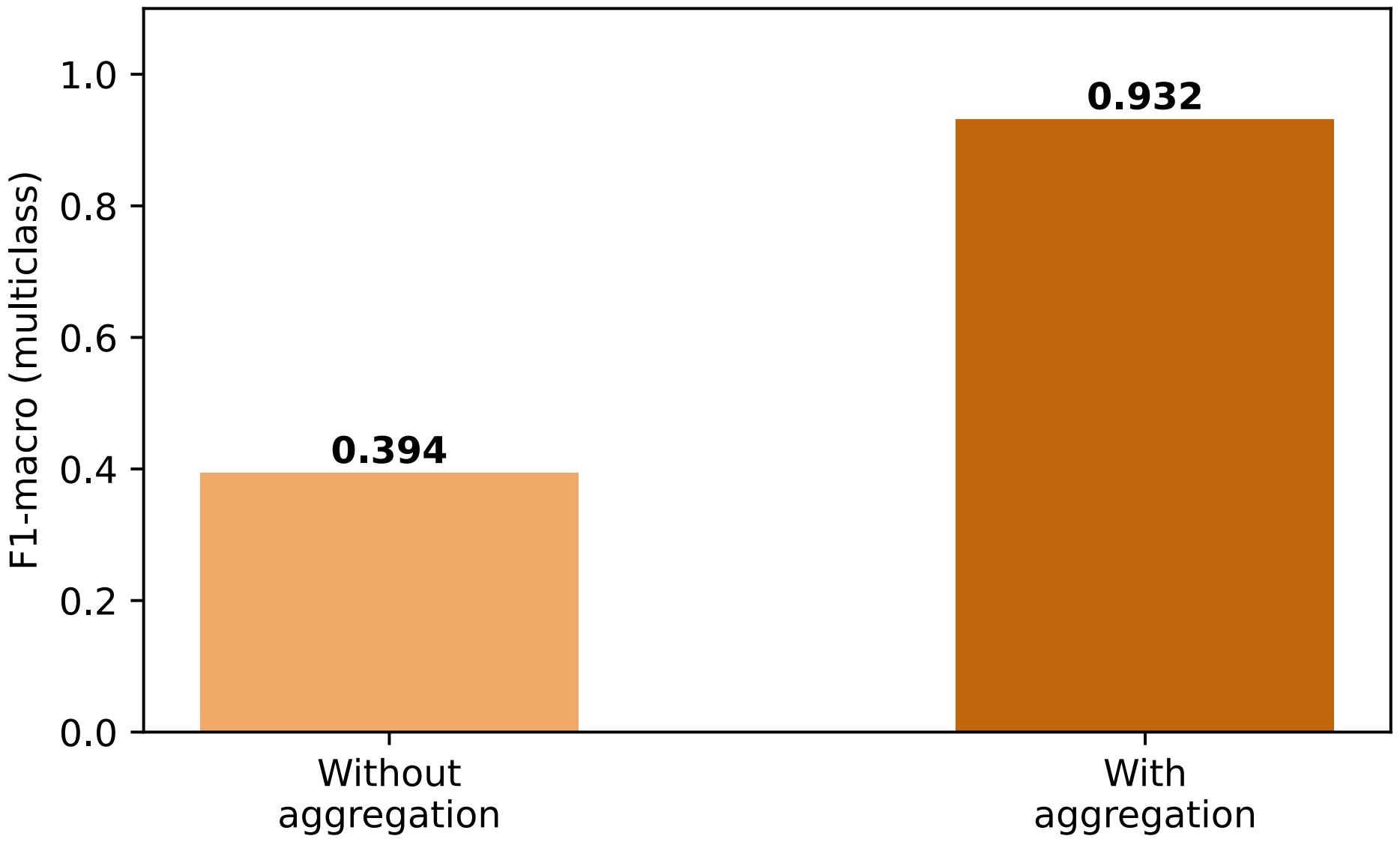

**Fig. 5.** Effect of causal temporal aggregation on multiclass F1-macro (Kitsune dataset).

An automated test confirms that the aggregation variable clearly distinguishes a genuine scan pattern (agg_distinct_dst_ports = 8) from repeated traffic to the same destination (agg_distinct_dst_ports = 1, n = 5). The performance effect is reported in Section 4.5: multiclass F1-macro on Kitsune rises from 0.394 to 0.932 (Fig. 5).

### *4.5 Final certified model results*

After incorporating aggregation, certified multiclass models were retrained across all three datasets, with a notable and consistent improvement, most pronounced on Kitsune (Table 3, Fig. 6).

**Table 3.** Certified multiclass model performance after incorporating causal flow aggregation.

| Dataset | F1-macro (with aggregation) | Classes | Training data origin |
|---|---|---|---|
| CICIDS2017 | 0.9538 | Normal, PortScan, DoS, BruteForce | Synthetic scenarios |
| UNSW-NB15 | 0.9491 | Normal, PortScan, DoS, BruteForce | Synthetic scenarios |
| Kitsune | 0.932 | Normal, PortScan, DoS, DDoS, Fuzzers, Generic, MitM (7) | Real data |

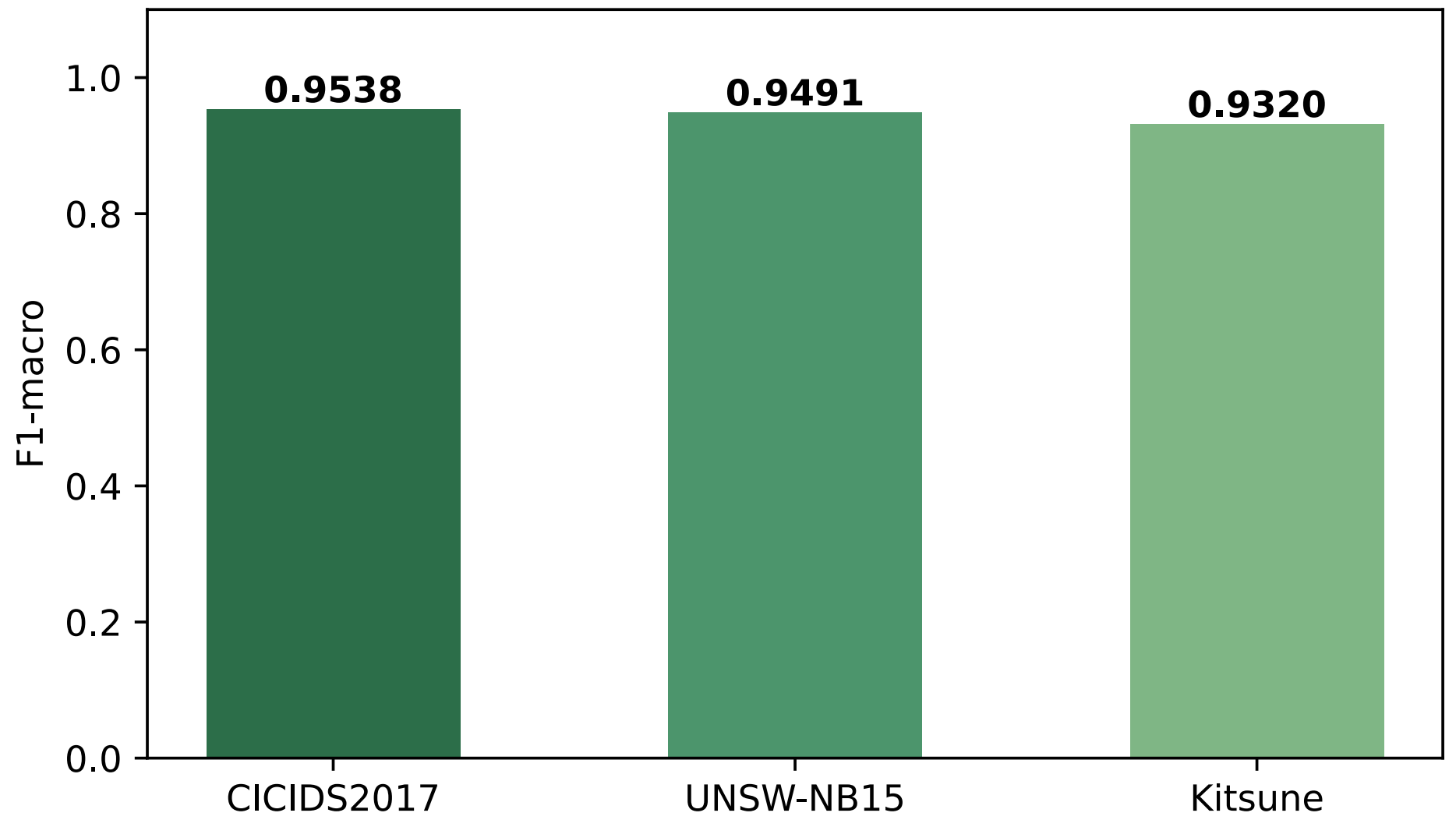

**Fig. 6.** Final certified multiclass model performance after incorporating causal flow aggregation.

The Kitsune case illustrates the aggregation effect directly: F1-macro rises from 0.3941 without aggregation to 0.9320 with it over 50 percentage points gained without modifying the XGBoost classifier itself, solely by enriching the feature vector with the origin's aggregated behavior. This confirms that much of the difficulty in correctly classifying certain attack categories, port scanning foremost, resided not in the classifier but in the absence of information about the temporal behavior of the traffic's origin.

By product-simplification decision, MODEXRE ships a single default certified model for Formal mode (model_v1, aggregation-enhanced, trained on CICIDS2017), since the universal ingestor normalizes all real evidence to the same representation regardless of origin. This default was later superseded by model_v3, the version distributed in the project repository, following the correction described in Section 5.2.

This default model (F1-macro = 0.9538, 4 classes) is preferred over the Kitsune-trained alternative (F1-macro = 0.9320, 7 classes) despite Kitsune's broader class coverage, because the choice prioritizes Normal-class performance, critical for minimizing false positives in a forensic context, and default-model consistency; the UNSW-NB15 and Kitsune aggregation-enhanced models remain certified and available in Laboratory mode for comparative analyses where Kitsune's finer-grained taxonomy is more relevant.

## 5. Explainability analysis and forensic case studies

### *5.1 Port-scanning case study*

Given that port scanning is the attack category this work's case study centers on, and that it is sufficiently represented across all three datasets, dedicated binary detectors (positive = PortScan, negative = any other case) were also trained using each source's reduced feature set (Table 4, Fig. 7).

**Table 4.** Specialized binary port-scan detector performance, by dataset.

| Dataset | Features used | F1-macro | Precision | Recall |
|---|---|---|---|---|
| CICIDS2017 | duration, packets_in/out, bytes_in/out | 0.9996 | 0.9997 | 0.9995 |
| UNSW-NB15 | duration, packets_in/out, bytes_in/out | 0.9224 | 0.9626 | 0.8902 |
| Kitsune | dport, sport | 0.9997 | 0.9999 | 0.9995 |

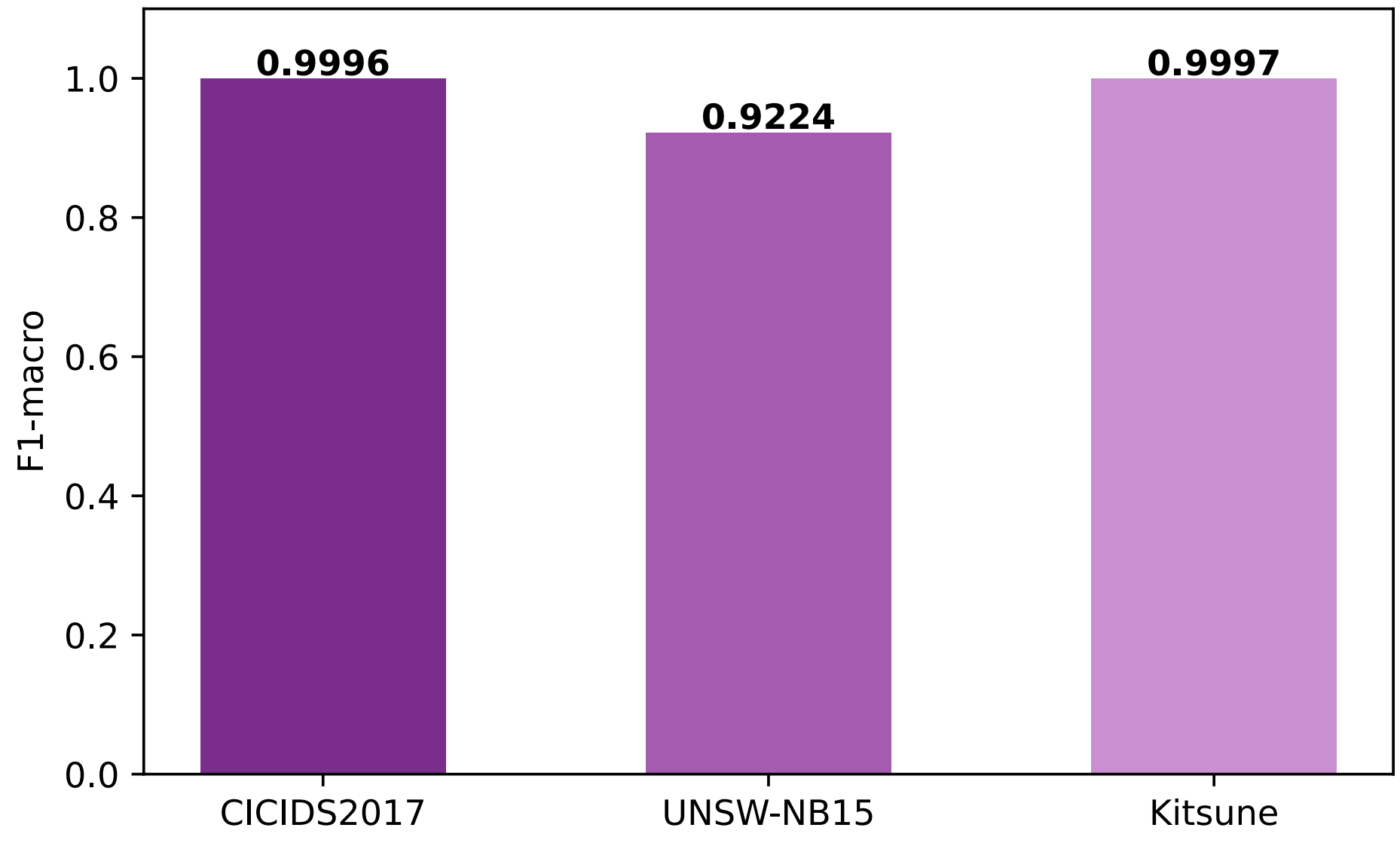

**Fig. 7.** Specialized binary port-scan detector performance, by dataset.

The Kitsune case is illustrative: while its generic multiclass F1-macro is 0.3941 (pre-aggregation; Section 4.5), the same pair of features (source and destination port) is nearly perfectly discriminative for the binary scan task (F1-macro = 0.9997). Destination-port diversity is an almost perfect signal for scanning but insufficient for the remaining six attack categories the same trade-off between informational richness and category granularity discussed in Section 4.4.

SHAP explanations are consistent with this analysis: in scan confirmations where aggregation is available, agg_distinct_dst_ports and agg_distinct_dst_hosts dominate the contribution to the prediction; without aggregation (e.g., the SSH alert of Section 4.3), bytes_in and duration dominate instead. This coherence between the SHAP-dominant variable and domain knowledge of each attack type is, per the qualitative evaluation criterion of Section 3.1, a necessary condition for the explanation to be forensically useful rather than a statistical artifact.

For the case-study port-scan event (Fig. 8), classified by the certified v1 model, agg_events_in_window, packets_in, and packets_out push the prediction toward PortScan, while duration and agg_distinct_dst_ports moderate it in the opposite direction for this specific event, a pattern consistent with a scanning host generating many brief, low-volume connections within the aggregation window.

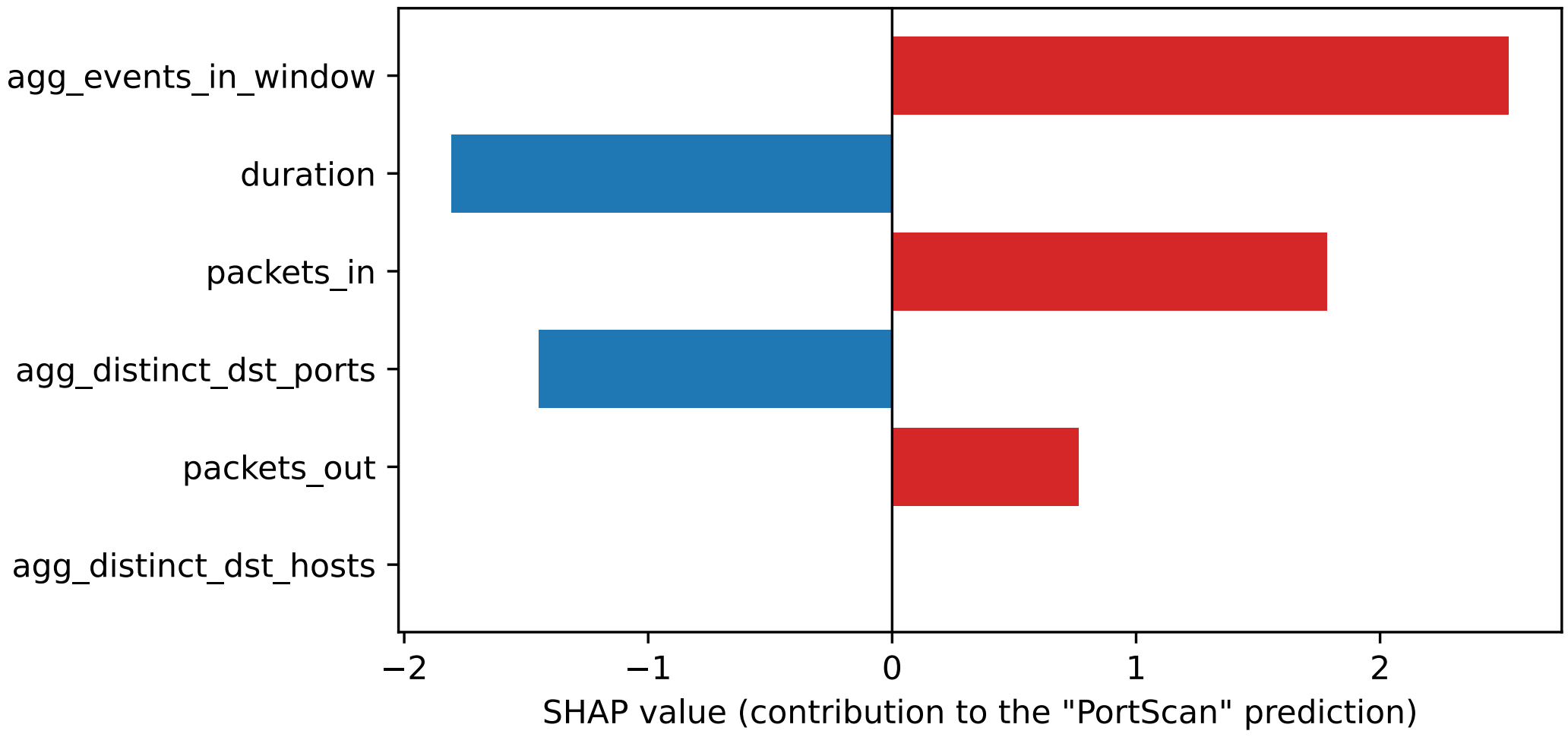

**Fig. 8.** SHAP attribution for the case-study port-scanning event, certified model v1.

### *5.2 Design decision 4: certification defect exposed by real evidence*

MODEXRE was subjected to a test not contemplated in earlier phases: processing a real packet capture (case_2026_001) of traffic from the authors' own host. The result was alarming: 100% of events (111 of 111) were classified as malicious (84 BruteForce, 27 PortScan), despite manual analysis confirming legitimate traffic (complete HTTP/HTTPS connections).

SHAP analysis revealed three independent causes. First, agg_events_in_window (a raw event count, not normalized by observation time) equated a host active during a short capture (16s, well below the 60s aggregation window) with a sustained attack; this was corrected with agg_events_per_second. Second, the closed taxonomy merged Reconnaissance into PortScan, with no dedicated category for horizontal sweeps. Third, and most consequential: the certification pipeline always trained a binary classifier from the label field, discarding attack_cat even when present in the source CSV no model had ever learned to distinguish attack categories.

All three causes were corrected and verified with dedicated regression tests: event-rate normalization, recovery of Reconnaissance and other native source-dataset categories (Bot, Worms, Heartbleed, Infiltration, Web Attack variants) in the closed taxonomy, and modification of the certification pipeline to train on attack_cat via multiclass encoding, with the class mapping persisted in the model manifest.

The artifact that embodies the defect, model_v2, was certified on 6 August 2026 on 242,361 training events and reports apparently excellent metrics (F1-macro = 0.9932, accuracy = 0.9969, macro precision = 0.9945, macro recall = 0.9919); yet its manifest records an empty class-name field. That absence is the fingerprint of the fault: the model had learned no attack taxonomy, only the binary distinction between normal and anomalous traffic that the certification flow silently imposed. It was therefore kept outside the public repository, which distributes only models v1 and v3.

The corrected model (model_v3) adds agg_events_per_second as a seventh feature and distinguishes 18 categories combining UNSW-NB15 and CICIDS2017, with F1-macro = 0.6214 (accuracy = 0.8908, macro precision = 0.7629, macro recall = 0.5889; n_train = 242,361, n_test = 60,591). The comparison with v2 isolates the effect of the correction: both models share the same seven features and the same training and test data and were certified forty-four minutes apart. F1-macro drops from 0.9932 to 0.6214 when the binary task is replaced by eighteen categories. This drop does not measure a worse system but the price of measuring the real task, which the earlier figure had concealed (Table 5). The gap between macro precision and macro recall describes a conservative model that errs more by omission than by wrongful attribution, the preferable failure mode in forensic practice, since an event left unclassified can be recovered in manual analysis, whereas a wrongly imputed attack category contaminates the report's conclusion.

**Table 5. Evolution of the certified MODEXRE models and the role of each.**

| Model | Features | F1-macro | Role in MODEXRE |
|---|---|---|---|
| Academic reference | 78 | 0.9992 | Comparison baseline; populates only 4 of 78 features on real evidence |
| Reduced coverage | 5 (duration, packets, bytes) | 0.9655 | First model applicable to real evidence (design decision 2) |
| Certified v1 | 6 (flow + 3 aggregation) | 0.9538 | Default Formal-mode model, 4 classes (design decision 3) |
| v2 (discarded) | 7 | 0.9932 | Artifact of the binary-only certification fault; empty class manifest |
| Certified v3 | 7 (v1 + event rate) | 0.6214 | Recertification of 18 categories; distributed in the repository |

The same capture, reanalyzed with model_v3, produced a radically different result: the large majority of events (about 87%) were classified as Normal, and the scanning host was correctly identified as Reconnaissance (mean probability 0.97), with agg_distinct_dst_hosts as the dominant SHAP variable, the same variable that, before correction, barely influenced the model's decisions.

This finding constitutes this work's clearest empirical validation: it does not rely on a laboratory test split, but on the same real case, before and after correction, with the cause identified and verified through automated tests. It is also the finding most directly relevant to forensic practitioners: a system with excellent benchmark metrics can nonetheless contain a defect invisible to every benchmark used during its development, detectable only by confronting it with evidence structurally different from that benchmark.

## 6. Discussion

### *6.1 Forensic implications*

The four design decisions reported in Sections 4 and 5 share a common pattern: none was anticipated at design time, and each was detected through a deliberate validation process (SHAP-based scrutiny of any suspiciously high or low model performance, or direct confrontation with real evidence) rather than through benchmark evaluation alone. This pattern has implications beyond this specific project: any team moving a machine learning model from a validated laboratory setting into a production environment with legal implications faces the same asymmetry between performance measured on a controlled test set and behavior against data that test set did not anticipate.

The methodological contribution of MODEXRE is therefore not to have avoided these failures through superior initial design, but to have built a development process capable of exposing and correcting them before the system was applied to a real case.

Beyond model drift, the dual-mode architecture has a second consequence: because Laboratory mode trains on public and synthetic data while private evidence is confined to Formal mode, certified models can be more powerful than case-specific ones and their training process can be shown in full before a court. This is arguably the design consequence with the widest reach outside this project.

Design decision 4 is the most representative instance of this pattern: a model with excellent validation-set metrics classified 100% of a real capture as malicious, without category discrimination, because the certification flow never trained on attack_cat. Only explicit validation against the same real case, before and after correction, made this detectable: no held-out split of any benchmark dataset used in this work would have revealed it.

### *6.2 Regulatory alignment*

MODEXRE's architecture operationalizes, rather than merely references, the admissibility requirements discussed in Section 2.3. Model-identity verification (Section 3.2) directly addresses the requirement that an AI system's output be tied to a specific, previously validated model. Chain-of-custody hash-chaining (Section 3.5) addresses evidentiary integrity requirements common to both the U.S. evidentiary framework and the EU AI Act's high-risk system obligations. Per-prediction SHAP explainability (Section 5) addresses the explainability requirement identified as a necessary not merely desirable condition for AI-generated evidence. These properties are reported here as measured system behavior (Sections 4 and 5), not as design aspirations, which is the distinction the admissibility literature draws between a system that claims compliance and one that demonstrates it.

### *6.3 Limitations*

Several limitations constrain the interpretation of these results without invalidating them.

First, the certified model has not yet been validated on the full-scale real dataset from the companion project; results rely on stratified samples sufficient to support the qualitative methodological conclusions but not to rule out that full-scale training would shift specific performance figures.

Second, evidence-source coverage is bound: the CEF parser covers the standard's common core rather than vendor-proprietary dialects, and PCAP flow aggregation is a purpose-built implementation of limited scope, not equivalent to a specialized tool such as CICFlowMeter. Replacing the zero-fill of absent features with native missing values (NaN), combined with retraining on simulated partial coverage, brought no improvement: SHAP analysis suggests a bias towards the majority Normal class in XGBoost's default direction for missing values, which aggravates rather than corrects the misclassification of evident cases with reduced coverage, so this line was discarded.

Third, the implemented chain of custody sits at an intermediate guaranteed level (hash-chain without digital signature): it detects alteration but does not support non-repudiable attribution of responsibility, a deliberate scope decision given this work's academic context.

Fourth, empirical validation has focused on port scanning and, following design decision 4, reconnaissance, while other taxonomy categories such as shellcode or backdoors have not received comparable depth of analysis.

Fifth, and most significantly for the regulatory alignment claimed in Section 6.2, MODEXRE has not been evaluated by legal practitioners nor subjected to an actual judicial process confirming that its explainability, traceability, and confrontability properties satisfy admissibility in practice; its regulatory alignment should be read as a literature-grounded design condition, not an empirically obtained procedural guarantee.

## 7. Conclusion

This work demonstrates that a laboratory-validated, explainable intrusion classifier can be moved into an operational forensic workflow with model-identity, evidence-normalization, and traceability guarantees that benchmark evaluation alone does not test, but that doing so requires a development process explicitly designed to expose the classifier's own failure modes before they reach a case report.

Four such failure modes were identified and corrected during this work: a dataset-construction artifact masquerading as signal, a coverage gap between academic reference features and real evidence, an architectural blindness to sequence-dependent attack patterns, and, most consequentially, a certification pipeline that had silently trained the wrong kind of classifier throughout development. The last of these was detectable only through direct confrontation with real evidence, not through any benchmark dataset used in this work, a finding with direct relevance beyond this specific system.

The final certified pipeline achieves F1-macro = 0.9538 on its default four-class model (v1), F1-macro = 0.6214 on the recertified 18-category model (v3) distributed in the repository, and up to 0.9997 on specialized binary port-scan detectors, with SHAP attributions consistent with domain knowledge of attack behavior and forensic KPIs (chain-of-custody integrity, model identity verification, and per-prediction explainability) reported as measured, auditable outcomes. MODEXRE remains a deliberately bound system, with limitations documented as rigorously as their positive results, and a modular architecture designed to accommodate the extensions those limitations suggest without requiring a redesign of its foundations. Future work includes digital signatures in the chain of custody (Level 2 to Level 3), model stacking to exploit the discriminative capacity of the 78-feature reference model on real evidence, extending validation to the remaining attack categories, evaluation by legal practitioners, and using MODEXRE as a validation framework for third-party AI systems.

## CRediT authorship contribution statement

**José Luis Vela:** Conceptualization, Methodology, Software, Investigation, Data curation, Writing - original draft, Visualization.

**Carmen Pellicer:** Conceptualization, Methodology, Writing - review & editing, Supervision, Project administration.

**Data availability**

The datasets used in this study are publicly available: CICIDS2017 (Canadian Institute for Cybersecurity, University of New Brunswick, https://www.unb.ca/cic/datasets/ids-2017.html), UNSW-NB15 (https://research.unsw.edu.au/projects/unsw-nb15-dataset), and Kitsune (https://archive.ics.uci.edu/dataset/516/kitsune+network+attack+dataset). **The MODEXRE source code, including the ingestion connectors, OCSF mapping layer, chain-of-custody implementation, and the 139-test regression suite, is available at https://github.com/PFM-Ceste/modexre.**

## Declaration of Generative AI and AI-assisted Technologies in the Writing Process

During the preparation of this manuscript, the authors used Claude (Anthropic) to assist with language editing, structural organization, and refinement of written sections. All content was critically reviewed, validated, and revised by the authors, who take full responsibility for the integrity and accuracy of the work reported. The AI tool was not used for data analysis, generation of results, or scientific decision-making.